\documentclass{article}
\usepackage{arxiv}
\usepackage{blindtext}
\usepackage{graphicx}
\usepackage{subfigure}
\usepackage{amsmath}
\usepackage{float}
\usepackage{booktabs}
\usepackage{authblk}
\begin{document}
\title{New method for coherent imaging using incompatible sources}
\author{Boyang Li}
\affil{Xi'an Institute of Optics and Precision Mechanics of Chinese Academy of Sciences, Xi'an, China 710119}
%\email{liboyang@opt.ac.cn}
%\author{Someone Else}

\maketitle
\begin{abstract}
Non-invasive imaging plays a crucial role in the early detection, diagnosis, and treatment of numerous medical conditions. This article discusses recent advancements in coherent imaging techniques.The article begins by discussing the evolution of CDI technology, focusing on its higher resolution capabilities for detailed analysis and the use of white light in modern systems. Subsequently, advancements in CDI techniques are examined, such as imaging for white matter tract analysis. And its applications in real-time visualization of internal structures is highlighted, along with its application to various samples. Finally introduce a method to use coherent imaging method on non-coherent sources.
\end{abstract}

\section{Introduction}

Since the first experimental demonstration of the coherent diffractive imaging (CDI) in 1999\cite{miao_extending_1999}, CDI has been widely utilized to biology \cite{Shapiro2005,Song2008b,Nishino2009,Nelson2010,Jiang2010,Giewekemeyer2010,Seibert2011,Kimura2014} and material\cite{Pfeifer2006,Chapman2006a,Song2008a,Robinson2009,Clark2013,Jiang2013} sciences.
With the emerging of more general light sources, the idea of capturing dynamics with complex illuminationwas brought to the table.

Coherent diffraction imaging (CDI) is an imaging technique that uses coherent beam to reconstruct images of micro and nanoscale objects. It is a powerful tool for imaging small scale structures with high resolution, and it has been used to image a wide range of materials, including proteins, viruses, and nanomaterials. Phase retrieval algorithms are used in coherent diffraction imaging (CDI) to reconstruct the phase of a diffracted wavefront from its intensity measurements. These algorithms are based on iterative optimization techniques, such as the Gerchberg-Saxton algorithm, the Hybrid Input-Output algorithm, and the Error Reduction algorithm. They are powerful and can recover images with high qualities. However most of the time, iterative phase retrieval algorithms suffers from twin images and stagnations. When the source becomes non-coherent. The CDI becomes hopeless. To overcome this problem, we report the new CDI method.

\section{Referenced pattern zooming}

Our method contains two steps.

The first step is referenced pattern zooming. Since most of the time, the reference is designed to be real and without fine structures, so it is easily reconstruced with conventional CDI algorithm such as HIO. In this step, we are able to measure slowly and precisely, meaning repeated measurement is possible to increase the dynamic range of the diffraction pattern and reduce noise. So we can have a precise, high quality reference known before the reconstruction of the sample.

\begin{equation}
  I_r(u,v) = |\mathcal{F}\{r(x,y)\}|^2
\end{equation}

The second step, both reference and sample are exposed to the beam. The pattern is recorded to reconstruct complex and fine structured sample. Using the information measured in the first step, by applying cross correlation constrain, fourier magnitude constrain and support constrain in sequence the image is reconstructed with high speed and high reconstruction rate. The algorithm is based on the following euqtions:

\begin{equation}
  h(x,y) = r(x,y) + o(x,y)\\
\end{equation}

\begin{equation}
  \begin{aligned}
    I_h(u,v) &= |\mathcal{F}\{h(x,y)\}|^2 \\
    &= |\mathcal{F}\{r(x,y)\}|^2 + |\mathcal{F}\{o(x,y)\}|^2\\
    &+\textbf{Re}\{\mathcal{F}\{r(x,y)\}(\mathcal{F}\{o(x,y)\})^{*}\}
  \end{aligned}
\end{equation}

\begin{equation}
  \begin{aligned}
    &\mathcal{F}\{I_h(u,v)-I_r(u,v)\} = o(x,y)\otimes o(x,y)\\
    &+r(x,y)\otimes o(x,y) + o(x,y)\otimes r(x,y)
  \end{aligned}
  \label{eq:autoc}
\end{equation}

\begin{equation}
  h(x,y) = r(x,y) + o(x,y)\\
\end{equation}

\begin{equation}
  \begin{aligned}
    I_h(u,v) &= |\mathcal{F}\{h(x,y)\}|^2 \\
    &= |\mathcal{F}\{r(x,y)\}|^2 + |\mathcal{F}\{o(x,y)\}|^2\\
    &+\textbf{Re}\{\mathcal{F}\{r(x,y)\}(\mathcal{F}\{o(x,y)\})^{*}\}
  \end{aligned}
\end{equation}

\begin{equation}
  \begin{aligned}
    &\mathcal{F}\{I_h(u,v)-I_r(u,v)\} = o(x,y)\otimes o(x,y)\\
    &+r(x,y)\otimes o(x,y) + o(x,y)\otimes r(x,y)
  \end{aligned}
  \label{eq:autoc}
\end{equation}

\begin{figure}[htbp]
\centering
\subfigure[] {\includegraphics[width=0.3\textwidth]{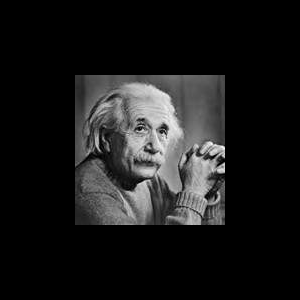}}
\subfigure[] {\includegraphics[width=0.3\textwidth]{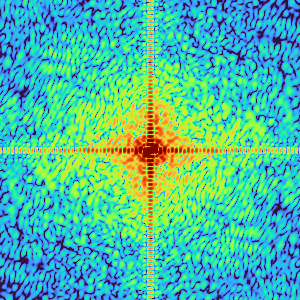}}
\subfigure[] {\includegraphics[width=0.3\textwidth]{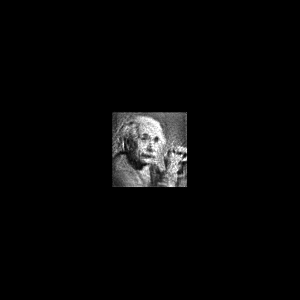}}
\caption{Demonstration for a CDI zooming.}
\label{fig:broad}
\end{figure}

\section{Referenced solver}

\begin{equation}
  \begin{aligned}
    &\mathrm{Find}~~o(x,y)\\
    &s.t.~(1)~I_h(u,v) = |\mathcal{F}\{o(x,y) + h(x,y)\}|^2\\
    &~~~~~(2)~xc(u,v) = 2\textbf{Re}\{\mathcal{F}\{r(x,y)\}(\mathcal{F}\{o(x,y)\})^{*}\}\\
    &~~~~~(3)~o(x,y) \in S
  \end{aligned}
  \label{eq:xcorr}
\end{equation}

Since now we have more constraints than usual, the uniqueness of the solution is enhanced. Simply using iterative projection is enough to recover the image:

During the reconstruction, the projection of (1) is:
\begin{equation}
  \begin{aligned}
    &H(u,v) = \mathcal{F}\{o(x,y)+r(x,y)\}\\
    &o(x,y) \leftarrow \mathcal{F}^{-1}\{\frac{\sqrt{I_h}}{|H(u,v)|}H(u,v)\}-r(x,y)\\
  \end{aligned}
\end{equation}

The projection of (2) is:
\begin{equation}
  \begin{gathered}
    O(u,v) = \mathcal{F}\{o(x,y)\},~~R(u,v) = \mathcal{F}\{r(x,y)\}\\
  \begin{aligned}
    O(u,v) \leftarrow& O(u,v)+\\
    &\frac{\frac{xc(u,v)}{2}-\textbf{Re}\{O(u,v)R(u,v)\}}{|R(u,v)|^2}R(u,v)\\
  \end{aligned}\\
    o(x,y) \leftarrow \mathcal{F}^{-1}\{O(u,v)\}\\
  \end{gathered}
\end{equation}

Lastly the projection of (3) is:
\begin{equation}
  \begin{aligned}
    &o(x,y) \leftarrow S(o(x,y))
  \end{aligned}
\end{equation}

\begin{figure}[htbp]
\centering
\subfigure[] {\includegraphics[width=0.3\textwidth]{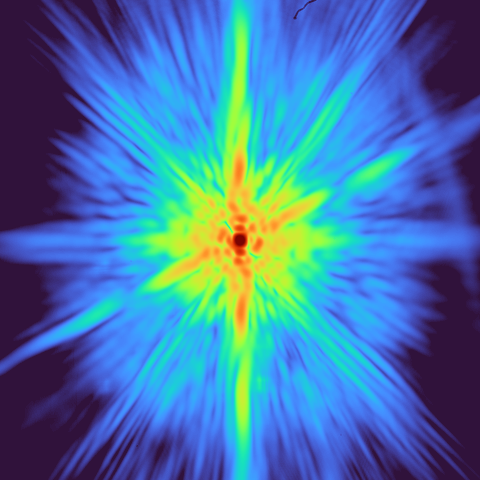}}
\subfigure[] {\includegraphics[width=0.3\textwidth]{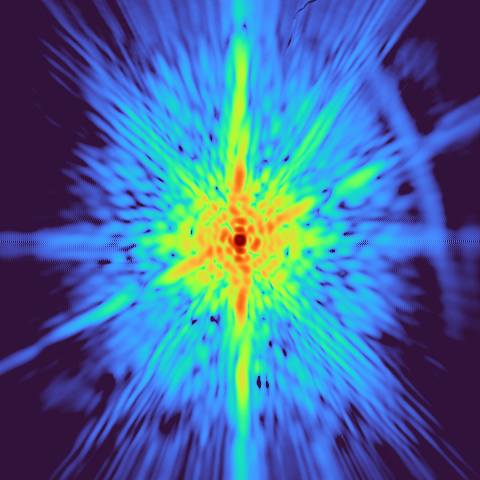}}
\subfigure[] {\includegraphics[width=0.3\textwidth]{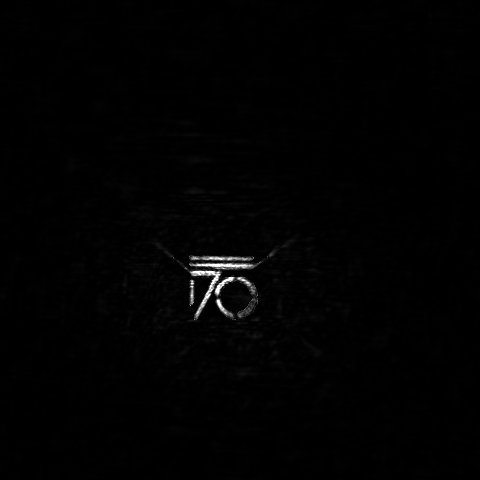}}
\caption{Algorithm testing using data from \cite{huijts_broadband_2020}.}
\label{fig:mono}
\end{figure}

The demonstration of our solver is shown in \ref{fig:mono}

This algorithm runs entirely on GPU.
It only took a few hours to get Fig. \ref{fig:mono}(b), which is super computation friendly.

\begin{figure}[htbp]
\centering
\subfigure[] {\includegraphics[width=0.24\textwidth]{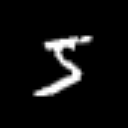}}
\subfigure[] {\includegraphics[width=0.24\textwidth]{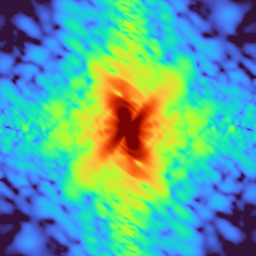}}
\subfigure[] {\includegraphics[width=0.24\textwidth]{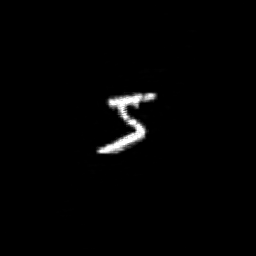}}
\caption{Algorithm testing using 3,5,7,9,11 harmonics.}
\label{fig:mnist}
\end{figure}

Another simulation is used to test the ability of this algorithm on discrete spectra.
We simulated diffraction of 3,5,7,9,11 harmonics of data from MNIST database extended as a $128\times128$ image as shown in Fig \ref{fig:mnist}(a) and (b).
Where the oversampling of 11 harmonics is 2.
The relative intensity is 0.2,0.4,0.4,0.3,0.2.
The reconstructed image with 1000 RAAR iterations and shrinking wrap each 20 iteration is shown in Fig \ref{fig:mnist}(d).

\begin{figure}[htbp]
\centering
\subfigure[] {\includegraphics[width=0.3\textwidth]{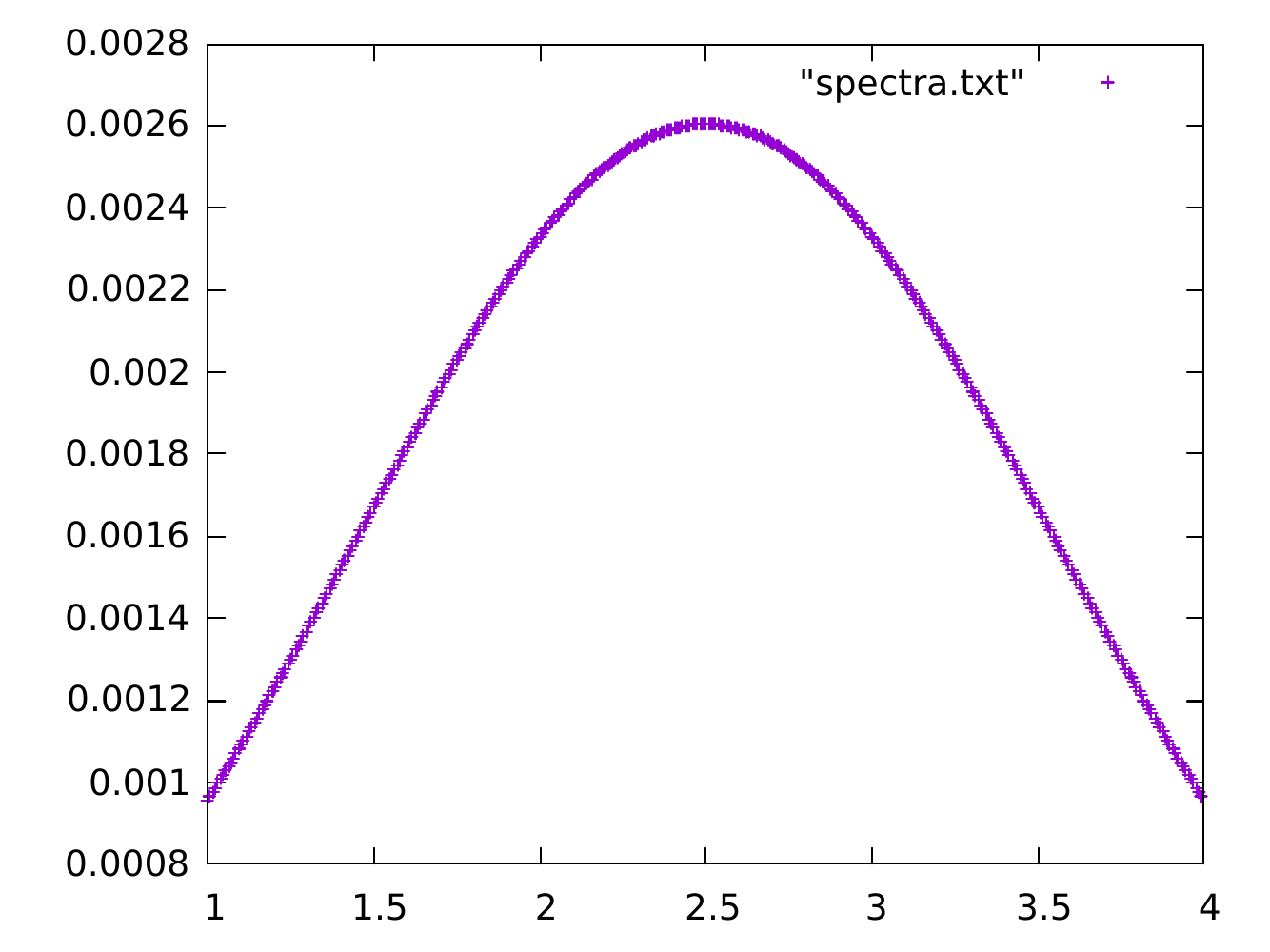}}
\subfigure[] {\includegraphics[width=0.22\textwidth]{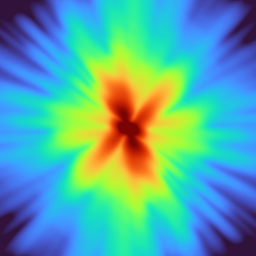}}
\subfigure[] {\includegraphics[width=0.22\textwidth]{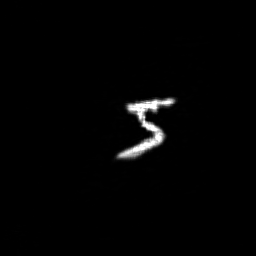}}
\caption{Algorithm testing using continuous spectrum with white light.}
\label{fig:mnistc}
\end{figure}

A continuous spectrum illumination is also simulated and tested.
The normalized spectrum is shown in Fig \ref{fig:mnistc} (a) with $\lambda_c=2.5$ and $\Delta\lambda/\lambda_c=80\%$.
384 points are taken for simulation.
The oversampling of $\lambda=1$ is 2. The same recipes as Fig. \ref{fig:mnist}(b-d) is shown in Fig \ref{fig:mnistc}(b-d).
More lambda component slows down the convergance. 500 iterations was used.

\bibliography{reference,cnncdi,insitu}

\bibliographystyle{unsrt}
\end{document}